\documentclass[superscriptaddress,aps, prd, reprint, nofootinbib]{revtex4-2}

\usepackage{graphicx}% Include figure files
\usepackage{amsmath,amssymb}
\usepackage{dcolumn}% Align table columns on decimal point
\usepackage{bm}% bold math
\usepackage{xcolor}
\usepackage{soul}

\newcommand{\ud}{\mathrm{d}}
\newcommand{\dd}{\mathop{\mathrm{d}\!}{}}
\newcommand{\lab}[1]{{\mathrm{#1}}}
\newcommand{\slab}[1]{{\textsc{#1}}}
\newcommand{\mb}[1]{{\mathbf{#1}}}

\newcommand{\sminus}{{\protect \scalebox {0.75}[0.7]{$-$}}}
\newcommand{\es}{\hspace{0.5pt}}
\newcommand{\floq}[1]{{\scalebox{0.65}{$(#1)$}}}

\renewcommand{\vec}[1]{\boldsymbol{\mathbf{#1}}}

\definecolor{blue2}{cmyk}{1, 0.1, 0.1, 0.1}

\newcommand{\GMT}[1]{\textcolor{orange}{#1}}

\usepackage{colortbl}
\definecolor{lightgreen}{cmyk}{0.2, 0, 0.2, 0.2}
\definecolor{lightgray2}{cmyk}{0.1,0.1,0,0.1}
\definecolor{Red2}{RGB}{214, 39, 40}
\definecolor{Blue2}{RGB} {31, 119, 180}
\definecolor{Orange2}{RGB}{255, 127, 14}
\definecolor{Green2}{RGB}{44, 160, 44}
\definecolor{greyish2}{rgb}{.96,.96,.96}

\definecolor{pyBlue}{RGB}{31, 119, 180}
\definecolor{pyRed}{RGB}{214, 39, 40}
\definecolor{pyGreen}{RGB}{44, 160, 44}
\definecolor{pyBlue2}{RGB}{0, 111, 237}
\definecolor{pyRed2}{RGB}{224, 52, 36}
\definecolor{Mathematica1}{rgb}{0.368417, 0.506779, 0.709798}
\definecolor{Mathematica2}{rgb}{0.880722, 0.611041, 0.142051}

\def\beq{\begin{equation}}
\def\eeq{\end{equation}}

\begin{document}

\title{Sharp Signals of Boson Clouds in Black Hole  Binary Inspirals}
% Force line breaks with \\

\author{Daniel Baumann}
\affiliation{Gravitation Astroparticle Physics Amsterdam (GRAPPA),
University of Amsterdam, Amsterdam, 1098 XH, Netherlands}
\affiliation{Center for Theoretical Physics, National Taiwan University, Taipei 10617, Taiwan}
\affiliation{Physics Division, National Center for Theoretical Sciences, Taipei 10617, Taiwan}
% \email{Author@institution.edu}

\author{Gianfranco Bertone}
\affiliation{Gravitation Astroparticle Physics Amsterdam (GRAPPA),
University of Amsterdam, Amsterdam, 1098 XH, Netherlands}

\author{John Stout}
\affiliation{Department of Physics, Harvard University, Cambridge, Massachusetts 02138, USA}

\author{Giovanni Maria Tomaselli}
\affiliation{Gravitation Astroparticle Physics Amsterdam (GRAPPA),
University of Amsterdam, Amsterdam, 1098 XH, Netherlands}

\begin{abstract}
Gravitational waves (GWs) are an exciting new probe of physics beyond the standard models of gravity  and particle physics. One interesting possibility is provided by the so-called ``gravitational atom,'' wherein a 
superradiant instability spontaneously forms a cloud of ultralight bosons around a rotating black hole. 
The presence of these boson clouds affects the dynamics of black hole binary inspirals and their associated GW signals.
In this Letter, we show that the binary companion can induce transitions between bound and unbound states of the cloud, effectively  ``ionizing'' it, analogous to the photoelectric effect in atomic physics. 
 The orbital energy lost in this process can overwhelm the losses due to GW emission, so that ionization drives the inspiral rather than merely perturbing it. We show that the ionization power contains sharp features that lead to distinctive ``kinks'' in the evolution of the emitted GW frequency. 
 These discontinuities are a unique signature of the boson cloud and observing them would not only constitute a detection of the ultralight boson itself, but also provide direct information about its mass and the state of the cloud.
\end{abstract}

%\keywords{Suggested keywords}%Use showkeys class option if keyword
                              %display desired
\maketitle

The dynamics of black hole mergers in vacuum is a precise prediction of general relativity (GR), which has been confirmed by the gravitational wave (GW) observations of the LIGO/Virgo Collaboration~\cite{PhysRevLett.116.061102, LIGOScientific:2021sio}. The robustness of these predictions implies that looking for any deviations is an interesting test for physics beyond the standard models of gravity and particle physics~\cite{Barack:2018yly,Bertone:2019irm}. Such signals would arise if new environmental effects modify the dynamics of the inspiral. For example, if dark matter clusters around black holes it would affect the inspiral through dynamical friction~\cite{Kavanagh:2020cfn, Coogan:2021uqv}.

Another interesting class of new physics, that can be probed with future GW observations, are \emph{ultralight bosons} with masses in the range of $10^{\sminus 20}$ -- $10^{\sminus 10}$ eV and very weak couplings to ordinary matter. 
Such weakly coupled particles arise in the string landscape as ultralight axions~\cite{Arvanitaki:2009fg, Mehta:2021pwf, Mehta:2020kwu,Demirtas:2018akl} and are also interesting dark matter candidates~\cite{Hui:2016ltb}.  However, these new hidden sectors do not necessarily have a large cosmic abundance, which makes detecting them an interesting challenge. 

Regardless of how these bosons couple to the Standard Model, however, they must gravitate. This universal coupling, plus the spate of current and upcoming gravitational wave detectors, has generated a lot of interest in using black holes, either in isolation or in an inspiral, to discover weakly coupled new physics~\cite{Barack:2018yly,Barausse:2020rsu,Bertone:2019irm,Chia:2020dye}. A particularly promising avenue relies on black hole superradiance~\cite{Brito:2015oca}, where a rapidly rotating black hole can spontaneously shed mass and angular momentum to form a large cloud of these ultralight bosons, independent of their prior cosmic abundance. 
The efficiency of this process depends on the ratio of the black hole's gravitational radius to the Compton wavelength of the field, the so-called gravitational fine structure constant,
\begin{equation}
    \alpha \equiv \frac{r_\lab{g}}{\lambda_\lab{c}} = \mu M\,, \label{eq:fineStructure}
\end{equation}
where $\mu$ is the boson mass, $M$ is the mass of the black hole and we use natural units, with $\hbar = c = G = 1$. 
For $\alpha \sim \mathcal{O}(0.01 - 0.1)$, the cloud both grows quickly and is long lived (on astrophysical timescales).

\begin{figure}[b!]
       \centering
       \includegraphics[width=0.95\columnwidth,trim={0pt 6pt 0pt 0}]{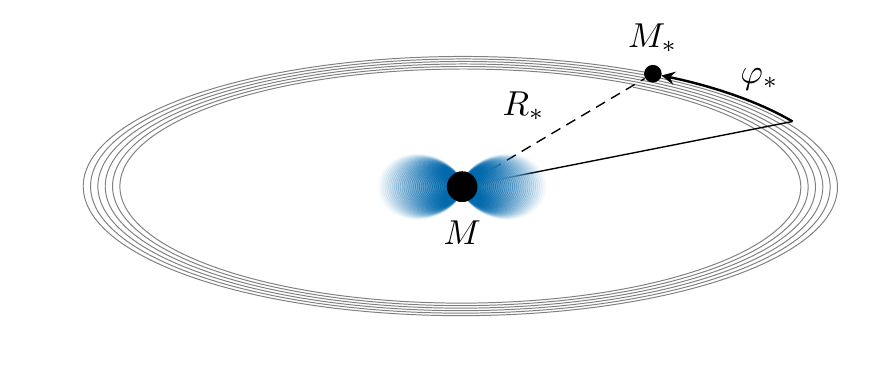}
       \caption{Schematic diagram of a gravitational atom in an equatorial binary inspiral. The position of the companion with mass~$M_*$ can be described by the distance between the two black holes, $R_*$, and the polar angle $\varphi_*$. }
        \label{fig:bplane} 
 \end{figure}

The structure of this boson cloud is nearly identical to that 
 a hydrogen atom and the system is often called a \emph{gravitational atom}. 
While superradiance naturally prepares the cloud in \emph{one} of these states, the others can be excited
when this atom participates in a binary inspiral (see Fig.~\ref{fig:bplane}).

The gravitational atom is, thus, subject to a quasi-periodic perturbation whose frequency slowly increases in time.  The strength of this nearly periodic  perturbation can be resonantly enhanced whenever its frequency matches the difference between two (or more) states of the cloud~\cite{Baumann:2018vus,Baumann:2019ztm}. In this Letter and its companion~\cite{Baumann:2021fkf}, we study bound-to-unbound state transitions of the gravitational atom, which allow part of the cloud to escape the gravity of the black hole. We refer to this process as ``ionization," in analogy to the photoelectric effect in atomic physics. The backreaction of the ionization process strongly
affects the dynamics of the inspiral.  As we will show, the departure from the GR predictions is significant, with the inspiral typically becoming much faster. This backreaction can be interpreted as dynamical friction acting on the body passing through the cloud~\cite{Zhang:2019eid,Traykova:2021dua, Annulli:2020ilw}.  Despite acting continuously as the separation shrinks, the effect contains sharp features carrying details on the energy structure of the cloud and the nature of the putative new particles. We will show how these sharp features are imprinted in the GW signals emitted by the binary.

\vskip 4pt
\paragraph*{\bf Gravitational atoms.} Consider a real scalar field~$\Phi$ of mass $\mu$, for simplicity without self-interactions. Using the ansatz $\Phi(t, \vec{r}) = \left[\psi(t, \vec{r}) e^{-i \mu t} + \text{h.c.}\right]/\sqrt{2\mu}$,
the Klein-Gordon equation  
in the Kerr background becomes an effective Schr\"odinger equation \cite{Dolan:2007mj,Baumann:2018vus}:
\begin{equation}
        i \frac{\partial}{\partial t} \psi(t, \vec{r}) = \left(-\frac{1}{2 \mu} \nabla^2 - \frac{\alpha}{r}+\cdots \right) \psi(t, \vec{r})\,, \label{eq:schrodinger}
    \end{equation}
where we have  neglected subleading terms in $\alpha$. 
The Schr\"odinger equation (\ref{eq:schrodinger}) has hydrogenic energy eigenstates, which can be divided into two qualitatively distinct classes.

The first are the \emph{bound states} $|n \es \ell \es m \rangle$, labeled by their total and azimuthal angular momentum $\ell$ and $m$, respectively, and the principal quantum number $n$. The bound state wave functions take the hydrogenic form, $\psi_{n\ell m}(t, \vec{r}\hskip 1pt)=R_{n\ell}(r)Y_{\ell m}(\theta,\phi)e^{\sminus i(\omega_{n\ell m}-\mu)t}$, where $\vec r \equiv (r, \theta, \phi)$ denote the Fermi frame which, at leading order in $\alpha$, coincides with the Boyer-Lindquist coordinates of the black hole. As for the hydrogen atom, the wave function varies over length scales set by the ``Bohr radius'' $r_\lab{c} \equiv (\mu \alpha)^{\sminus 1}$ and  decays exponentially as $r \to \infty$. For small $\alpha$, the frequencies are~\cite{Detweiler:1980uk,Dolan:2007mj,Baumann:2019eav} 
\begin{equation}
        \omega_{n \ell m} = \mu\!\es\es \left(1 - \frac{\alpha^2}{2 n^2} + \mathcal{O}\big(\alpha^4\big)\right) + i\Gamma_{n\ell m}\,, \label{eq:boundEnergies}
\end{equation}
where the imaginary part $\Gamma_{n \ell m} \propto \mu \alpha^{4 \ell + 5}$ comes from
the dissipative nature of the black hole's event horizon. This imaginary part permits the superradiant growth or decay of the bound states, depending on whether they rotate faster or slower than the horizon. We work in a range of parameter space in which these growth rates are fast enough to ensure the growth of the cloud on astrophysical timescales, yet are slow enough that they can be ignored when focusing the timescales relevant to the binary inspiral. In this approximation, the bound states are described by a discrete set of hydrogenic energies $\epsilon_{n \ell m} = \lab{Re}\es\es [\omega_{n \ell m}] - \mu$. By convention, these states are unit-normalized $\langle n \es \ell \es m | n' \es \ell' \es m' \rangle = \delta_{n n'} \delta_{\ell \ell'} \delta_{m m'}$, where the inner product is defined in the standard way.

The second class of eigenstates are the \emph{unbound states} $|\epsilon; \ell m \rangle$, which are labeled by the non-negative energy $\epsilon = \omega - \mu \geq 0$, total angular momentum $\ell$, and azimuthal angular momentum $m$. In contrast to the bound states, the unbound wave functions $\psi_{\epsilon; \ell\es m}(t, \vec{r}) = R_{\epsilon; \ell}(r) Y_{\ell m}(\theta, \phi)$ have purely real frequencies and asymptote to spherical waves with wavenumber $k$, such that $\epsilon = k^2/(2 \mu)$, as $r \to \infty$. By convention, these unbound states are normalized such that $\langle \epsilon; \ell\es m | \epsilon'; \ell' \es m'\rangle = \delta(\epsilon - \epsilon') \delta_{\ell \ell'} \delta_{m m'}$\, and are orthogonal to every bound state.

\vskip 4pt     
\paragraph*{\bf Ionization.} An inspiraling companion with mass $M_* \equiv q M$, orbital separation $R_*$, and true anomaly $\varphi_*$ (see Fig.~\ref{fig:bplane}) perturbs the Schr\"{o}dinger equation (\ref{eq:schrodinger}) by its gravitational potential,
\begin{equation} 
            V_*(t,\vec{r}\hskip 1pt) =  -q \alpha \sum_{\ell, m} \frac{r_{\scriptscriptstyle \! < }^{\ell}}{r_{\scriptscriptstyle >}^{\ell+1}} \,  \varepsilon_{\ell m}\, e^{\sminus i m \varphi_*} \es  Y_{\ell m}  (\theta, \phi)   \, , \label{eq:companionPerturbation}
\end{equation}
where the sum\footnote{We explicitly exclude both the $\ell = 0$ and $\ell = 1$ contributions, as the latter is fictitious~\cite{Baumann:2018vus} and neither mediate level transitions.} ranges over $\ell \geq 2$ and $|m| \leq \ell$, and $r_{\scriptscriptstyle >}$ ($r_{\scriptscriptstyle < }$) denotes the larger (smaller) of $r$ and $R_*$. We restrict to inspirals that take place entirely within the equatorial plane, so that the \emph{tidal moments} simplify\GMT{:} $\varepsilon_{\ell m} = \frac{4 \pi}{2 \ell + 1} Y_{\ell m}^*(\frac{\pi}{2}, 0)$, where the superscript $*$ denotes complex conjugation.
The gravitational perturbation (\ref{eq:companionPerturbation}) is \emph{quasi-periodic}, with both the orbital frequency\footnote{By convention, the positive (negative) sign denotes an orbit in which the companion co-rotates (counter-rotates) with the cloud.} $\Omega(t) = \pm \ud\varphi_*/\ud t$ and the orbital separation $R_*(t)$ slowly evolving as the parent black hole and companion merge. For simplicity, we consider quasi-circular orbits, for which $\ud \Omega/\ud t = \gamma (\Omega/\Omega_0)^{11/3}$, where the \emph{chirp rate} $\gamma$ is defined with respect to the reference frequency $\Omega_0$ as $\gamma = \frac{96}{5} q M^{5/3} \Omega_0^{11/3}/(1 + q)^{1/3}$. Throughout the inspiral phase, however, this frequency evolves very slowly $\gamma \ll \Omega^2$ which allows us to linearize the frequency $\Omega(t) \approx \Omega_0 + \gamma t$ in the regime we are interested in. The cloud is, thus, subject to a gravitational perturbation whose fundamental frequency $\Omega(t)$ slowly increases in time. 

In this Letter, we study how the cloud and binary inspiral evolve when this driving frequency is high enough to efficiently mediate transitions from the bound to unbound states of the atom, resonantly unbinding or \emph{ionizing} the cloud from the black hole. We first assume that the system initially occupies a single bound state $|n_\lab{b} \es \ell_\lab{b} \es m_\lab{b} \rangle$ whose energy we denote by~$\epsilon_\lab{b}$. The companion (\ref{eq:companionPerturbation}) connects this state to the continuum of unbound states $|\epsilon; \ell\es m \rangle$ 
via a matrix element with definite frequency $\langle \epsilon; \ell\es m | V_*(t) | n_\lab{b} \es \ell_\lab{b} \es m_\lab{b} \rangle = \eta_{\ell m}(\epsilon; t)\exp [-i(m- m_\lab{b})\varphi_*(t)]$, where the amplitude $\eta_{\ell m}(\epsilon; t)$ inherits its slow time dependence from the companion's radial motion.

As described more thoroughly in Ref.~\cite{Baumann:2021fkf}, at weak couplings $q \alpha \ll 1$, the evolution of the system can be divided into three stages, centered about the time $t_0$ where the frequency of the perturbation matches the minimum energy difference between the initial state and the continuum, $\pm (m - m_\lab{b}) \Omega(t_0) + \epsilon_\lab{b} = 0$. Far before this time, $\sqrt{\gamma}(t - t_0) \ll -1$, the perturbation oscillates too slowly to provide enough energy to excite the bound state into the continuum, and so the cloud remains bound. Far after this time, $\sqrt{\gamma}(t - t_0) \gg 1$, the companion orbits quickly enough to ionize the cloud and the cloud steadily depletes. For intermediate times, $\sqrt{\gamma} | t - t_0| \lesssim 1$, there are transient phenomena which interpolate between these two regimes over a timescale that is fast compared to that of the inspiral and set by~$\gamma^{\sminus 1/2}$.

The steady depletion of the cloud is well approximated by applying Fermi's Golden Rule, which states the transition rate per unit phase space (energy) into unbound states with angular quantum numbers $\ell$ and $m$ is
\begin{equation}
    \ud \Gamma_{\ell m} = \ud \epsilon \, \big|\eta_{\ell m}(\epsilon; t)\big|^2\, \delta\big( \epsilon - \epsilon_b \mp (m - m_\lab{b})\es  \Omega(t)\big)\,. \label{eq:fermiGolden}
\end{equation}
While Eq.~(\ref{eq:fermiGolden}) is typically derived by assuming that neither $\Omega(t)$ nor $\eta_{\ell m}(\epsilon; t)$ evolve in time, it is consistent~\cite{Baumann:2021fkf} to apply it here as long as $q \alpha \ll 1$ and we ignore any transient phenomena. Summing this rate over \emph{all} unbound states that the cloud can transition into yields an equation for the mass ejected from the cloud by ionization,
\begin{equation}
    \left.\frac{\ud M_\lab{c}}{\ud t}\right|_{\lab{ion}} = \,- M_\lab{c}\sum_{\ell, m} \big|\eta_{\ell m}\big(\epsilon_{\smash{*}}^{\floq{m}}; t\big)\big|^2\, \Theta\big(\epsilon_{\smash{*}}^\floq{m}\big)\,,
    \label{eqn:IonizationRate}
\end{equation}
where $\epsilon_{\smash{*}}^\floq{m}(t) = \epsilon_b + g \es\es \Omega(t)$ is the energy of the state the cloud ionizes into and $\Theta(\epsilon)$ is the Heaviside step function. We also define $g \equiv \pm(m - m_\lab{b})$, where the positive (negative) sign denotes co-rotating (counter-rotating) orbits.

\begin{figure}
      \centering
      \includegraphics[width=1\columnwidth,trim={0 6pt 0pt 0}]{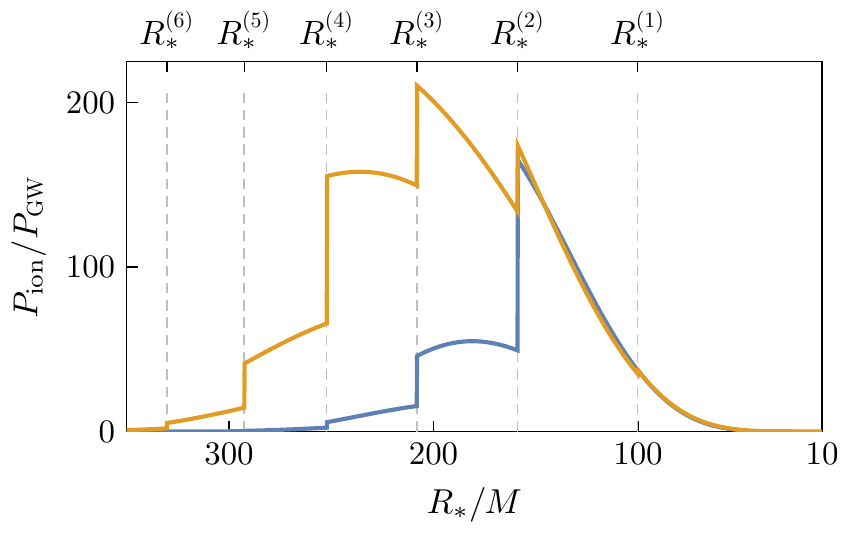}
      \caption{Ratio between the energy (per unit time) lost due to ionization and due to GW emission for  co-rotating [{\color{Mathematica1} blue}] and counter-rotating [{\color{Mathematica2} orange}] orbits, with $\alpha=0.2$ and $M_{\lab{c}}/M=0.01$  in a $|211\rangle$ state. }
      \label{fig:ionizationPower}
    \end{figure}

The companion must do work to ionize the cloud. The \emph{ionization power}, i.e.~the energy lost by the orbit per unit time due to ionization, is
\begin{equation}
    P_\lab{ion} = \frac{M_\lab{c}(t)}{\mu}\sum_{\ell, m} g\es\es \Omega(t)\es \big|\eta_{\ell m}\big(\epsilon_{\smash{*}}^\floq{m}; t\big)\big|^2\, \Theta\big(\epsilon_{\smash{*}}^{\floq{m}}\big)\,.
     \label{eqn:IonizationPower}
\end{equation}
In Fig.~\ref{fig:ionizationPower}, we compare this ionization power to the energy lost due to GW emission,
\begin{equation}
    P_\slab{gw} = \frac{32}{5} \frac{q^2 M^2}{(1 + q)^2} R_*^4\Omega^6\,,
\end{equation}
ignoring cloud depletion for both co- and counter-rotating orbits, as a function of the binary separation~$R_*$.
We see that ionization is a \emph{large} effect, as it takes energy from the binary orders of magnitude more efficiently than GWs, even with a conservative choice of the cloud's initial mass~$M_{\lab{c},0}$. 

\vskip 4pt
\paragraph*{\bf Sharp features.} The most distinctive feature in Fig.~\ref{fig:ionizationPower} are the \emph{sharp} jumps in the ionization power at specific orbital separations 
\begin{equation}
      \frac{R_{\smash{*}}^{\floq{g}}}{M} = \alpha^{\sminus 2} \left[4 g^2 (1+q) n_\lab{b}^4\right]^{1/3}\,,{\quad\,\, g = 1, 2, \es\es \cdots\,.}
      \label{eqn:discontinuities}
\end{equation}
These jumps arise in co-rotating (counter-rotating) orbits because, for each set of unbound states with azimuthal angular momentum $m > m_\lab{b}$ ($m < m_\lab{b}$), there is a frequency at which $\epsilon_{\smash{*}}^\floq{m} = 0$ and the companion can just begin to ionize the bound state $|n_\lab{b} \es \ell_\lab{b} \es m_\lab{b} \rangle$ into that continuum. The values of $R_{\smash{*}}^\floq{g}$ in Eq.~(\ref{eqn:discontinuities}) correspond to the orbital separations at those frequencies.

These jumps appear sharp because $|\eta_{\ell m}(\epsilon; t)|^2$ is \emph{finite} in the low-energy limit, $\epsilon \to 0$. This is due to the long-ranged nature of the gravitational potential, which localizes the low-energy unbound states about the black hole,
\begin{equation}
    \lim_{\epsilon \to 0} R_{\epsilon; \ell}(r) = \sqrt{\frac{2 \mu}{r}} J_{2 \ell + 1}\big( 2 \sqrt{2 \mu \alpha r}\big),
\end{equation}
and forces them to have non-vanishing transition elements to the bound states~\cite{Baumann:2021fkf}. The aforementioned transient phenomena that occur near the moments when~$\epsilon_{\smash{*}}^{\floq{m}}(t) = 0$ soften this sharp behavior. 

    \vskip 4pt
    \paragraph*{\bf Scaling symmetry.} In the limit of small $q$, the ionization power (\ref{eqn:IonizationPower})  has an interesting scaling symmetry which allows us to determine the result for arbitrary parameters after it has been computed once for a fiducial set of parameters.
The radial wave functions $R_{n\ell}(r)$ and $R_{\epsilon;\ell}(r)$ depend only on the dimensionless variables $r/r_\lab{c}=\alpha^2r/M$ and $k\es r=\sqrt{2\mu\epsilon}\es r$, respectively. Moreover, the energy $\epsilon_{\smash{*}}^{\floq{m}}$ appearing in Eqs.~(\ref{eqn:IonizationRate}) and (\ref{eqn:IonizationPower}) scales as $\alpha^3/M$ times a function of $r/r_\lab{c}$. This means that, when evaluating the matrix elements $\eta_{\ell m}(\epsilon; t)$ at $\epsilon_{\smash{*}}^{\floq{m}}$, all the radial variables will scale as $\alpha^2r/M$.
     The ionization power must then also scale homogeneously with~$\alpha$. By power counting, we find that 
\begin{equation}
    P_\lab{ion} =\alpha^5 q^2 (M_\lab{c}/M) \, \mathcal{P}(\alpha^2 R_*/M)\, , \label{eqn:scaling-Pion}
\end{equation}
where $\mathcal P$ is a universal function for each $|n_\lab{b} \es \ell_\lab{b} \es m_\lab{b} \rangle$ that can be computed numerically.

\vskip 4pt
\paragraph*{\bf Accretion.} 
If the companion is also a black hole, it will absorb some of the cloud as it 
passes through it. 
The capture cross section of an ultralight scalar field by
a black hole has been computed in Refs.~\cite{Unruh:1976fm,Benone:2014qaa,Benone:2019all}. As the companion moves through the bulk of the cloud, $R_* \sim r_\lab{c}$, it will accrete mass at a rate~\cite{Baumann:2021fkf}
\begin{equation}
\frac{\dd M_*}{\dd t}=\mathcal{A}_*\es\rho \,,
\end{equation}
with $\mathcal A_* = 16\pi M_*^2$ the horizon area of the companion, whose rotation we ignore for simplicity, and $\rho$ the mass density of the medium. 
In addition to a sizable increase in its mass which enhances $P_\lab{ion}$ and $P_\slab{gw}$, the companion also experiences a significant force as it absorbs momentum from the cloud.

\vskip 4pt
\paragraph*{\bf Binary evolution.} To determine the impact of ionization and accretion on the binary inspiral, we numerically solve for the evolution of a few benchmark systems.
 We will focus on intermediate mass ratio binaries, with $q \ll 1$, for which the sharp features of the ionization can occur in band for a future space-based observatory like LISA. We solve for the 
 separation $R_*$, %the mass ratio $q \equiv M_*/M$ and the cloud--black hole mass ratio $q_\lab{c}\equiv M_\lab{c}/M$. 
 the companion's mass $M_*$ and the cloud's mass $M_\lab{c}$.
As resonances between bound states happen only at specific discrete orbits \cite{Baumann:2019ztm}, while ionization and accretion act continuously, we will ignore the former.
Our solutions, therefore, should not be taken as a complete study of the system's dynamics, but rather as demonstrations of the impact that ionization and accretion can have on an inspiral.

The evolution of $M_*$ and $M_\lab{c}$ is governed by mass conservation,
\begin{align}
\frac{\dd M_*}{\dd t} &= 16\pi M_*^2 \rho(\mb{R}_*)\,, \label{equ:q-evolve} \\[4pt]
\frac{\dd M_\lab{c}}{\dd t} &= - \frac{\dd M_*}{\dd t}- M_\lab{c}\sum_{\ell, m} \big|\eta_{\ell m}\big(\epsilon_{\smash{*}}^{\floq{m}}; t\big)\big|^2\, \Theta\big(\epsilon_{\smash{*}}^\floq{m}\big)\,, \label{equ:qc-evolve}
\end{align}
where $\rho(\mb{R}_*) \equiv M_\lab{c}|\psi|^2$ is the local density of the cloud at the position of the companion.  
For a real field~$\Phi$, the cloud is not axisymmetric, and, thus, we replace $\rho(\mb{R}_*)$ with its average over an orbit. The second term on the right-hand side of Eq.~(\ref{equ:qc-evolve}) captures the mass loss due to ionization as defined in Eq.~(\ref{eqn:IonizationRate}). The evolution of 
$R_*$ follows from the conservation of energy, 
\begin{equation}
\begin{aligned}
    \frac{qM^2}{2R_*^2}\frac{\dd R_*}{\dd t}={}&{-P_\slab{gw}}-P_\lab{ion} \\[-4pt]
    &-
    \biggl(\sqrt{MR_*}\mp\frac{mM}{\alpha}\biggr) \biggl(\frac{M}{R_*} \bigg)^{3/2}\frac{\dd q}{\dd t}\,,
    \end{aligned}
    \label{eqn:evolution-R}
\end{equation}
where we have neglected terms that are subleading in~$q$. The right-hand side of Eq.~(\ref{eqn:evolution-R}) includes the radiation reaction force from the emission of GWs and the friction caused by both ionization and accretion.

As illustrated in Fig.~\ref{fig:ionizationPower},
$P_\lab{ion}$ can overwhelm $P_\slab{gw}$ for a wide range of separations. The evolution of the binary will then be \emph{driven}, rather than simply perturbed, by the interaction with the cloud.
The extra friction can dramatically shorten the merger time and, generally, 
 a ``plunge'' is observed as soon as $P_\lab{ion}$ overcomes~$P_\slab{gw}$.

\vskip 4pt
\paragraph*{\bf Imprints in the GW signal.}  The binary emits gravitational waves with frequency \mbox{$f_\slab{gw}=\Omega/\pi$}, where \mbox{$\Omega^2=M/R_*^3$} for a circular Keplerian orbit. Distinct features in $R_*(t)$, therefore, become observable signatures 
in~$f_\slab{gw}(t)$ and, hence, the observed GW waveform. 

\begin{figure}[t]
            \centering
            \includegraphics[trim={0 6pt 0pt 0}]{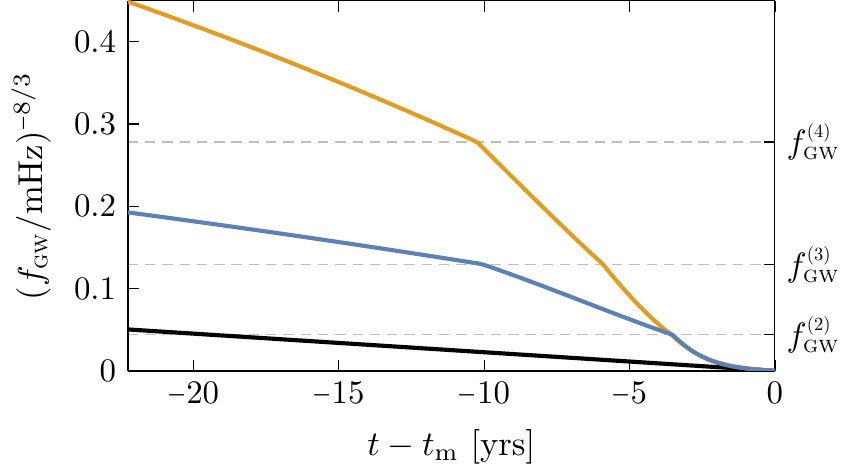}
            \caption{Evolution of the GW frequency as a function of the time to merger, $t-t_\lab{m}$, for $M=10^4M_\odot$ and $\alpha=0.2$, with initial values of $R_{*}=400M$, $q =10^{-3}$ and $M_{\lab{c}}/M=0.001$ in a $|211\rangle$ state. Shown are the results for co-rotating [{\color{Mathematica1} blue}] and counter-rotating [{\color{Mathematica2} orange}] orbits, relative to the vacuum solution [black].}
    \label{fig:evolution-frequency}
\end{figure}

Figure~\ref{fig:evolution-frequency} shows the evolution of the GW frequency for co- and counter-rotating orbits, with a power-law rescaling that transforms the vacuum solution into a straight line. 
It is immediately apparent that the deviations from the vacuum solution are both \emph{large} and feature distinctive \emph{kinks}, which arise from the discontinuities in~$P_\lab{ion}(R_*)$ at the separations (\ref{eqn:discontinuities}).  
These kinks are a unique signature  of the boson cloud and carry significant information about
 the parameters of the system.  From (\ref{eqn:discontinuities}), the GW frequency at the kinks is
   \begin{equation}
   \begin{aligned}
      f_\slab{gw}^{(g)} &= \frac{6.45\,\text{mHz}}{g}\left(\frac{10^4M_\odot}{ M}\right)\!\left(\frac{\vphantom{10^{4} M_\odot}\alpha}{0.2 \vphantom{M}}\right)^{\!3}\!\left(\frac{2}{n_\lab{b}}\right)^{\!2} \\
      &=\frac{33.5 \,\text{mHz}}{g} \left(\frac{M}{10^4M_\odot}\right)^{\!2}\!\left(\frac{\vphantom{M}\mu}{10^{\sminus 14}\,{\rm eV}}\right)^{\!3}\!\left(\frac{2}{n_\lab{b}}\right)^{\!2}\! .
      \end{aligned}
      \label{eqn:f-discontinuities}
    \end{equation}
In Fig.~\ref{fig:evolution-frequency}, we have chosen parameters such that
     these kinks occur in the range probed by future space-based GW detectors like LISA. 
    This requires relatively large $
    \alpha$ for which the lifetime of the $|211\rangle$ state becomes a concern~\cite{Baumann:2019ztm,Baumann:2021fkf}. To account for this decay, we choose a small initial value for the mass of the cloud, $M_\lab{c}/M = 10^{\sminus 3}$.  
  It is possible to have smaller values of $\alpha$ if we simultaneously reduce $M$, although the degree to which this is possible is limited by the fact that \mbox{$q = M_*/M$} must be small enough for our perturbative analysis to be valid.   Measuring kinks at specific frequencies tells us about the state of the cloud and the mass of the field $\mu$, especially if the black hole mass $M$ can be measured with other parts of the signal. 

\begin{figure}[t!]
            \centering
            \includegraphics[width=0.48\textwidth,trim={0 6pt 0pt 0}]{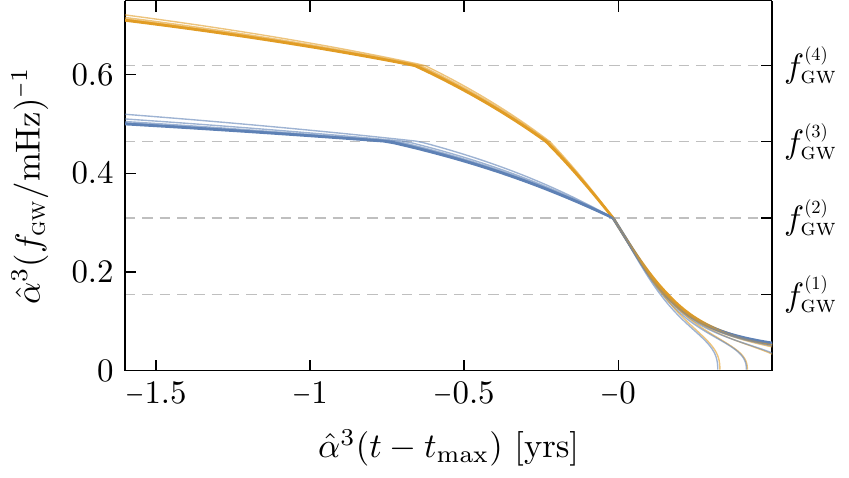}
            \caption{Evolution of the (inverse) frequency $f_\slab{gw}$ for $M=10^4M_\odot$ and $\alpha=0.04,0.08,\ldots,0.28$, with initial $q=10^{\sminus 3}$ and $M_{\lab{c}}/M=0.01$ in a $|211\rangle$ state. The axes are rescaled according to Eq.~(\ref{eqn:solution-scaled-f(t)}),  with $\hat \alpha \equiv \alpha/0.2$. The curves have been horizontally shifted to match at $t=t_\lab{max}$, which has been chosen close to the peak of $P_\lab{ion}/P_\slab{gw}$. Shown are the results for co-rotating [{\color{Mathematica1} blue}] and counter-rotating [{\color{Mathematica2} orange}] orbits.}
    \label{fig:frequency-scaling}
\end{figure}

Figure~\ref{fig:evolution-frequency} presented the evolution of the system for a specific choice of parameters. 
In the regime of interest, $P_\lab{ion} \gg P_\slab{gw}$, the dependence on these parameters can be determined analytically using a scaling symmetry of the evolution equations.
Neglecting 
other forces and changes in $q$ and $M_\lab{c}$ throughout the inspiral, we can obtain an approximate equation for the evolution of $f_\slab{gw}$ under the effect of ionization only:
\beq
\frac{\dd f_\slab{gw}^{2/3}}{\dd t} \approx \frac{2}{\pi^{2/3}} \frac{P_\lab{ion}}{q M^{5/3}}\, .
\eeq
Using the ionization power's scaling behavior (\ref{eqn:scaling-Pion}), 
we can write this as 
\beq
\frac{\dd z^{2/3}}{\dd \tau} \approx 2 \hskip 1pt \mathcal P(z^{\sminus2/3})\,, 
\eeq
where we have defined the dimensionless variables $z \equiv (M/\alpha^3) \hskip 1pt \pi f_\slab{gw}$ and $\tau \equiv \alpha^3qM_\lab{c} t/M^2$.  The solution can, therefore, be written as
\beq
f_\slab{gw}(t) = \frac{\alpha^3}{M} f(\tau(t))\, ,
\label{eqn:solution-scaled-f(t)}
\eeq
where $f(\tau)$ is a universal function that depends on the shape of $P_\lab{ion}$ for a given state $|n_\lab{b}\es \ell_\lab{b} \es m_\lab{b}\rangle$ of the cloud. The region of validity of this formula increases with larger $M_\lab{c}$. In Fig.~\ref{fig:frequency-scaling}, we confirm that the solutions of the full system of Eqs.~(\ref{equ:q-evolve}--\ref{eqn:evolution-R}) indeed are described by a universal shape, when both $f_\slab{gw}$ and $t$ are appropriately rescaled. The curves depart from each other only when the approximation $P_\lab{ion}\gg P_\slab{gw}$ fails (that is, very close and very far from the merger) or when corrections due to the varying $q$ and $q_\lab{c}$ become important.

\vskip 4pt
\paragraph*{\bf Conclusions.}

The dynamical impact of superradiant clouds on binary inspirals is phenomenologically rich. In this Letter, we studied a new effect---ionization---that is important when the binary separation is comparable to the size of the cloud. The orbital energy lost in the process can overwhelm the losses due to GW emission, so that ionization drives the inspiral rather than just perturbing it. Although it acts continuously throughout the inspiral, ionization also leaves sharp, distinct signatures in the frequency evolution of the system and carries direct information on the state of the cloud and the mass of the scalar field. 

Our analysis made a number of simplifying assumptions. Most notably we neglected the resonant bound-to-bound transitions mediated by the gravity of the companion. Their inclusion is necessary to understand the history of the system and the evolution of the state of the cloud.  Moreover, while we restricted ourselves to quasi-circular equatorial orbits, interesting effects could arise in the general case, such as orbital plane precession or eccentrification. A combined treatment of all of these effects will serve as a starting point to model gravitational waveforms involving gravitational atoms and devise suitable strategies to discover them with upcoming GW detectors.

\vskip 4pt
\paragraph*{\it Acknowledgements}
D.B.~and J.S.~are grateful to Horng Sheng Chia and Rafael Porto for previous collaborations on this topic.
D.B.~receives funding from a VIDI grant of the Netherlands Organisation for Scientific Research~(NWO) and is part of the Delta-ITP consortium.  
D.B.~is also supported by a Yushan Professorship at National Taiwan University funded by the Ministry of Science and Technology (Taiwan). J.S.~is supported by NASA Grant No.~\texttt{80NSSC20K0506}.

\newpage
\bibliography{main2}% Produces the bibliography via BibTeX.

\end{document}